**Gate Tunable In- and Out-Of-Plane Spin-Orbit Coupling and Spin Splitting Anisotropy at LaAlO₃/SrTiO₃ (110) Interface**

By *Kalon Gopinadhan\*, A. Annadi, Younghyun Kim, Amar Srivastava, Brijesh Kumar, Jingsheng CHEN, J. M. D. Coey, Ariando, and T. Venkatesan\**

[*]    Dr. Kalon Gopinadhan, Dr. A. Annadi, Amar Srivastava, Dr. Brijesh Kumar,
        Prof. Jingsheng CHEN, Prof. Ariando, Prof. T. Venkatesan
NUSNNI-NanoCore, National University of Singapore, Singapore 117576
E-mail: gopinadhan@iitkalumni.org (KG);eletv@nus.edu.sg (TV)
        Amar Srivastava, Prof. Ariando
Department of Physics, National University of Singapore, Singapore 117542
        Prof. T. Venkatesan
Department of Electrical and Computer Engineering, National University of Singapore, Singapore 117576
        Dr. Kalon Gopinadhan, Prof. Jingsheng CHEN, Prof. T. Venkatesan
Department of Materials Science and Engineering, National University of Singapore, Singapore 117575, Singapore
        Younghyun Kim
Department of Physics, University of California, Santa Barbara, California 93106,USA
        Prof. J. M. D. Coey
School of Physics and CRANN, Trinity College, Dublin 2, Ireland

**Manipulating spin-orbit coupling (SOC) is important for devices such as spin-orbit torque based memory and its understanding is neccessary to answer several fundamental open questions in triplet state superconductivity, topological insulators and Majorana fermions. Here we report spin splitting of 25 meV at the LaAlO₃/SrTiO₃ (110) interface for in-plane spins at a current density of $1.4 \times 10^4$ A/cm², which is large compared to that found in semiconductor heterostructures or the LaAlO₃/SrTiO₃ (100) interface, and in addition it is anisotropic. The anisotropy arises from the difference in electron effective mass along the [001] and [1-10] directions. Our study predicts a spin splitting energy > 1000 meV at a current density of $10^7$ A/cm², which is enormous compared to metallic systems and will be an ideal spin polarized source. In addition to**

the in-plane effect, there is an unexpected gate-tunable out-of-plane SOC at the $LaAlO_3/SrTiO_3$ (110) interface when the spins lie out-of-plane due to broken symmetry in the plane of the interface. We demonstrate that this can be manipulated by varying the $LaAlO_3$ thickness showing that this interface can be engineered for spin-orbit torque devices.

## 1. Introduction

Spin-orbit coupling (SOC) – the interaction of an electron's spin with its motion – is the basis of many of the most interesting phenomena in magnetism. It is vital to contemporary topics in condensed matter physics such as triplet superconductivity,[1] topological insulators[2] and Majorana fermions.[3] It offers the prospect of manipulating spin states by the application of an electric field, for example, in devices such as the proposed Datta-Das field-effect spin transistor.[4] Recently, spin-orbit torque induced magnetization switching in soft magnetic materials has been actively investigated for magnetic memory applications (as this requires much smaller current densities compared to direct current switching of magnetic domains) and a strong SOC in a layer adjacent to the soft magnetic layer is key to switching in-plane[5] or perpendicular magnetization.[6] Spin-orbit coupling in atoms is usually significant only in high-Z elements because the interaction increases as the fourth power of the atomic number. However, disruption of the periodic boundary conditions in a solid, particularly broken inversion symmetry at surfaces and interfaces, also leads to SOC. The classic examples are interfaces of GaAs/GaAlAs[7] and InAs/InGaAs[8] and more recently $LaAlO_3/SrTiO_3$ (100) (LAO/STO).[9-11] Recently, it has been shown that at the (001) surface of $SrTiO_3$, the Ti $3d$ orbital subband is spin-split by 90 meV with spins of opposite chirality, using spin- and angle-resolved photoemission spectroscopy.[12]

A large perpendicular electric field is expected at the LAO/STO interface due to broken structural inversion symmetry which might produce a Rashba SOC. In the Rashba effect, the conduction electrons of wave vector $k$ moving in a potential gradient normal to the interface $(\partial V / \partial z)$, experience an effective in-plane magnetic field $B_{so}$ where $B_{so} = -\frac{1}{m^{*}c}(\hbar k \times \partial V / \partial z)$ . The Rashba interaction is represented by the Rashba Hamiltonian $H = \alpha \sigma . (\hat{z} \times \hbar k)$ where $\sigma$ is the Pauli spin matrix, $\hat{z}$ is a unit vector normal to the interface and $\alpha$ is the spin-orbit coupling strength. In conventional two-dimensional Rashba systems, the spin direction is always in-plane which is useful for out-of-plane magnetization switching, however here we show that the LAO/STO (110) interface is a good candidate for spin-orbit torque based magnetic switching using both in-plane and out-of-plane spin-orbit interactions, where the terms 'in-plane and 'out-of-plane' describe the spin directions. We find that that the magnitudes of these two contributions depends upon the LAO thickness.

The discovery of a gate tunable two-dimensional conductivity at the LAO/STO interface associated with unusual properties such as magnetism,[13,14] superconductivity,[15-17] and Kondo scattering[18] has spurred intense research activity over the last decade. Most studies were based on (100) LAO/STO interfaces but two-dimensional conductivity has also been reported at the (110) LAO/STO interface[19,20] which is significant because the crystallographic orientation influences the electrical properties. An in-plane Rashba SOC has been reported at the (100) interface,[10,11] but there is no report yet on the (110) interface. Moreover, the (110) interface provides an opportunity to understand the effect of crystalline anisotropy, strain and band-filling on the SOC which could help us engineer the interfaces for spin-orbit torque induced magnetization switching.

In this study, we report the first out-of-plane gate tunable SOC in a (110) system. In addition we report the anisotropic in-plane SOC at the LAO/STO (110) interface. We find that the strength of the SOC is affected by interface strain via the thicknesses of LAO overlayer, which results in z-confinement and modifications in band-filling. We also show that the crystal orientation affects SOC via the band structure.

## 2. Results and Discussion

Voltage gated oxide interfaces are attractive as it allows one to tune the chemical potential electrostatically. In a single band scenario, one would expect that for negative gate voltages ($V_G$) on the STO substrate, the electrons are pushed away from the interface resulting in a decrease in carrier density ($n_e$), whereas in the positive $V_G$, electrons are attracted to the interface resulting in an increase in $n_e$. However in reality, gate voltage dependence of $n_e$ can be much more complex since conduction band of STO at the interface has contributions from multiple-orbitals due to z-confinement and tetragonal distortion in contrast to semiconductors. To investigate the nature of band filling in STO (110), we have studied the effect of back gating on the transport parameters. The schematic diagram of the measurement geometry is shown in **Figure 1**(a) where the current flows either in the [001] or [1-10]  direction. A back gate voltage ($V_G$) has been applied across 0.5 mm thick STO substrate. An applied $V_G$ of 10 V will correspond to an electric field of $2x10^2$ V/cm. Schematic of the Ti-terminated STO (110) surface with the Ti and O atoms arrangement along [001] and [1-10] crystallographic directions is shown in Figure 1(b). In the [001] direction the 180° Ti-O-Ti bonds are undistorted (lower electron mass) whereas in the [1-10] direction there are zig-zag 90° Ti-O-Ti bonds with less overlap (larger electron mass). Figure 1(c) shows Hall coefficient $R_H$ ($V_G$) in both crystallographic directions [001] and [1-10] for 5 uc of LAO at 1.9 K. The magnitude and gate tunability of $R_H$ is similar in both directions. We notice that at a critical $V_G$ of ~ -5 V, $R_H$ decreases abruptly. It is reported that for 2DEG based on STO (100), there appears a universal carrier density of $1.68 \times 10^{13}$ cm$^{-2}$ at a critical gate voltage above which multiband conduction occurs due to a transition from a symmetric $d_{xy}$ orbital to less symmetric $d_{xz}/d_{yz}$ orbitals[21] and the mobility is expected to either remain constant (if only light mass carriers are probed via Hall measurement ) or to decrease (if both light and heavy mass carriers are probed) above the critical carrier density. However, in our samples for $V_G > -5$ V, an increase in mobility is seen which argues against multi-band conduction as we discuss below. From the observed linear Hall effect up to a magnetic field of 9 Tesla, the presence of multiple carriers is not evident (see supplementary details). Figure 1(d) shows the variation of mobility with gate voltage along [001] and [1-10]. Here the mobility increases monotonically as $V_G$ is increased above the critical voltage, and the behavior is similar in both directions. This can be understood in terms of the overlap of the 2p oxygen orbitals with the 3d orbitals of Ti which determines the transmission amplitude and hence the conductivity. Near the interface, oxygen octahedra ($TiO_6$) tilt in order to accommodate the strain at the interface arising from the lattice mismatch between LAO (a = 3.745 Å) and STO (a = 3.905 Å) resulting in a lower electron transmission across the orbitals.[22]

We shall now discuss the variation of $R_H$ with $V_G$ on the 15 uc sample which is shown in Figure 1(e). The $R_H$ shows significant anisotropy, in contrast to the 5 uc sample. Along [1-10], $R_H$ initially decreases with $V_G$ and at a critical $V_G \sim 20$ V, it increases. The variation of $R_H$ along [1-10] is similar to what Joshua *et al.* [21] have reported for high mobility samples of LAO/STO (001) interface. On the other hand along [001], $R_H$ increases with increase in $V_G$. This behavior is similar to what Dikin *et al.* [13] have reported for LAO/STO (001), and a multi-band conduction is proposed as the most probable explanation. Significantly, above the critical voltage, the anisotropy in $R_H$ between the two directions is negligible. This unusual result suggests that the electric field effect on oxide interface is quite different from the conventional semiconductor interfaces. Indeed, the subband filling needs to be considered. Figure 1(f) shows the mobility along [001] and [1-10] and it shows a small anisotropy only above the critical voltage which is similar to 5 uc sample, however the increase in mobility is much steeper here. In addition, at $V_G = 0$ V, the mobility is 60 times smaller than 5 uc of LAO sample which means that the oxygen octahedra at the interface are tilted far more than for the 5 uc due to increased number of LAO layers which exert a bigger compressive strain on the STO surface. The 2DEG is ~ 7 nm below the surface.[23] An applied $V_G$ relaxes the distortion[24] and as a result the mobility increases and the rate of increase in mobility is much higher than 5 uc of LAO due to a larger distortion in the 15 uc case. This result also implies that as LAO thickness increases, in addition to the band filling effects, strain at the interface also changes which is modified by an applied $V_G$. A systematic decrease in mobility of two orders of magnitude has been reported for LAO/STO (001) interface upon increase in thickness of LAO from 5 to 25 uc[25] and the possible reason is a modification in the tilt of the oxygen octahedra near the interface.[26]

To understand the nature of SOC at this interface, the magnetoconductance has been measured at a temperature of 1.9 K. The measured magnetoconductance ($\Delta\sigma(H)$) along [001] for a sample with 5 uc of LAO for various values of the back gate voltage $V_G$ is shown in Figure 2(a) when the magnetic field H is applied normal to the interface. The negative magnetoconductance with a cusp at low magnetic fields is a signature of weak antilocalization of the carriers due to SOC.[27,28] In addition, we have independently verified the presence of SOC and magnetic ordering in the 2DEG through anisotropic magnetoresistance (AMR) measurements.[9] To extract the SOC parameters $\tau_{so}$, the spin relaxation time and $\tau_\phi$, the inelastic relaxation time, the magnetoconductance is analyzed using the Kim-Lutchyn-Nayak (KLN) theory,[29] assuming a Rashba SOC linear in electron wave vector (see supplementary) along with a Zeeman coupling term which gives a very good fit to the experimental results. The weak antilocalization measurements in each direction is directly due to the interference between electron waves travelling in close loops in a diffusive medium which is still two-dimensional in nature. Since there is no existing Rashba SOC theory for (110) interface, we analyzed our results based on isotropic (001) Rashba theory (KLN) which should in principle yield similar results, however the anisotropy would be smaller due to averaging effects. In addition, we have performed fitting to Maekawa-Fukuyama (MF) model[30] (results of MF fitting are given in supplementary details). From both fits, we find identical SOC parameters, however the KLN theory gives a better fit up to high magnetic fields (~9 T).

We also fitted the magnetoconductance assuming a Rashba SOC cubic electron wave vector, (shown in Figure 2(a) and also see supplementary) and it is clear from the data that we do not see any significant difference between linear and cubic Rashba fits. The KLN theory yields a good fit to the data for $V_G < 0$, however it is poor for $V_G > 0$, which could be due to the presence of multiple-conduction channels such as $d_{xy}$ and $d_{yz}/d_{xz}$ orbitals and Zeeman effect. Therefore we restrict our analysis and discussion to the $V_G < 0$ case. The magnetoconductance for the 5 uc sample along [1-10] is shown in Figure 2(b) which is similar to [001], with a slightly larger negative conductance for the same gate voltage. In this case the KLN fits deviate from the data at high magnetic fields. To know the effect of LAO thickness on the magneto-transport, we measured the magnetoconductance for the 15 uc LAO sample and the results are shown in Figure 2(c) and 2(d). The low field magnetoconductance in this case is isotropic with respect to the crystal orientation and significantly different from the 5 uc case. The thickness of LAO has a critical influence on the transport parameters of 2DEG. In the 15 uc LAO case also, it is hard to distinguish linear and cubic Rashba SOCs as either fits the data well. It is to be noted that in the 15 uc case, the carrier density $n_e$ varies from $1.8 \times 10^{14}$ to $5 \times 10^{13}$ cm$^{-2}$ with gate voltage, whereas for 5 uc sample, $n_e$ varies only from $2.2 \times 10^{13}$ to $2.5 \times 10^{13}$ cm$^{-2}$. Thus, the two samples cover a fairly large range of carrier densities, which helps us to probe the nature of SOC at the interface.

To determine the dynamics of the SOC, the spin relaxation time $\tau_{so}$ is extracted from the fitting parameter $q_{so}$ as defined in KLN theory[29] using $q_{so}^2 = 1/D\tau_{so}$ where $D$ is the diffusion coefficient which is estimated from the Einstein relation, that is $D = \pi \hbar^2 n_e \mu/(m^* e)$. Here $n_e$ is the sheet carrier density and $\mu$ is the mobility both of which were estimated from the Hall measurements. The effective mass of the electron in both crystallographic orientations is taken as $m^* = 3m_0$ [31] as the first approximation, where $m_0$ is the free electron mass. On the other hand, the inelastic phase coherence time $\tau_\phi$ is calculated using the fitting parameter $q_\phi^2 = 1/D\tau_\phi$. KLN theory has $q_{so}, q_\phi$ and $q_z$ (which is related to the Zeeman effect) as the parameters for fitting the magnetoconductance data. The magnitude of $\tau_{so}$ is different along [001] and [1-10], suggesting an anisotropic spin relaxation mechanism for the LAO/STO (110) 2DEG. The spin relaxation mechanism is best understood from the dependence of $\tau_{so}$ with elastic scattering time $\tau$ (or $\mu$) where $\tau = m^* \mu / e$: a linear relation indicates Elliott-Yafet mechanism[32] and an inverse relation indicates D'yakonov- Perel' (Rashba) mechanism.[33] **Figure 3**a shows the variation in $\tau_{so}$ as a function of $1/\mu$ along [001] and [1-10] for the 5 uc LAO sample. In both directions, there is an inverse relationship between $\tau_{so}$ and $\mu$, which implies a Rashba SOC at the interface. Figure 3(b) shows the inverse relation between $\tau_{so}$ and $\mu$ for the 15 uc LAO sample. Here also, the sample exhibits an inverse relationship between $\tau_{so}$ and μ.

We now calculate the strength of the SOC, $\alpha$ which is related to the fitted inverse spin relaxation length $q_{so}$ through $\alpha = \hbar^2 q_{so} / 2m^*$. Figure 3(c) shows the variation in $\alpha$ as a function of $V_G$ for the 5 uc sample. The evolution of $\alpha$ with $V_G$ has two linear regions of different slopes suggests that there could be changes in the orbital occupancy or an orbital splitting with up and down spins.[34] In addition, the anisotropy of $\alpha$ along [001] and [1-10] suggests that the effective mass of the conduction electrons is higher along [1-10] than [001] since $\alpha$ is inversely proportional to the effective mass. Naively, the electron movement along [1-10] requires a zig-zag path between Ti orbitals (Figure 1(a)) which decreases the transmission amplitude and hence a larger effective mass for the conduction electron in contrast to a larger transmission amplitude along [001]. A recent angle resolved photoemission (ARPES) study shows that the band structure of STO(110) based surface is very different from STO(001) and is highly anisotropic.[35] In the presence of broken inversion symmetry, the dispersion relationship at the Fermi surface exhibits a splitting in the k-vectors for up and down spins. The spin-splitting energy is given by $\Delta = 2\alpha k_F$ where $k_F$ is the Fermi wave vector which is related to carrier density as $k_F = \sqrt{2\pi n_e}$. At $V_G = -10$ V, $\Delta$ is estimated to be 25 meV and 15 meV along [001] and [1-10] respectively, which is much higher than that of 5.5 meV reported for InGaAs/InAlAs heterostructures,[36] 1.6 meV for InAlAs/InGaAs/InAlAs quantum wells,[37] 1.4 meV for InAs quantum wells,[38] and 10 meV for LAO/STO (100) heterostructures.[10] The larger $\Delta$ for (110) interface as compared to (100) interface is inferred to be a result of the enhanced degree of structural inversion asymmetry[37] [7] which increases $\alpha$ and hence $\Delta$. It should be noted that the current through 10 μm wide patterned structure in our sample is 1 μA, which implies a current density of 1.4 x$10^4$ A/cm$^2$ assuming a 2DEG thickness of 7 nm as reported previously.[23] Recently, we reported a spin-orbit field at the LAO/STO (100) interface that is proportional to current.[9] This suggests that the spin splitting energy can be > 1000 meV for a reasonable current density of $10^7$ A/cm$^2$. The change in carrier concentration is small for the range of the applied $V_G$. Thus $\Delta \propto \alpha$ suggesting that the spin-splitting is also anisotropic. Figure 3(d) shows the variation in $\alpha$ as a function of $V_G$ for the 15 uc sample. Here the $V_G$ dependence of $\alpha$ shows smaller anisotropy than the 5 uc sample, and it can be explained by the additional contribution of strain in 15 uc sample (as evident from the poor mobility) leading to a stronger quantum confinement. This creates a larger potential gradient ($\partial V / \partial z$) making the in-plane crystallographic orientation dependence irrelevant.

To understand the difference between the two samples, especially the strain and z-confinement effects, angle dependent magnetoconductance measurements were performed. **Figure 4**(a) and 4(b) shows the data obtained at $V_G$ = -40 V for 5 and 15 uc of LAO by varying the angle ($\phi$) of the applied magnetic field to the interface, from out-of-plane ($\phi$ = 0°) to in-plane ($\phi$ = 90°) direction. As the magnetic field orientation changes from out-of-plane to in-plane, the negative magnetoconductance decreases much more for the 5 uc sample than for the 15 uc sample. This means that the z-confinement of carriers should be stronger in the case of 15 uc of LAO. The splitting of Ti t$_{2g}$ orbitals is known to be a combined result of the quantum confinement and tetragonal distortion of $TiO_6$,[12,39] which means that the position and occupation of $d_{xy}$, $d_{yz}/d_{zx}$ orbitals in the energy band diagram for 15 uc sample should be very different from the 5 uc sample. Overall, these results suggest that the carrier confinement and orbital filling play a significant role in oxide interface physics.

Estimation of inelastic dephasing time ($\tau_\phi$) and spin-orbit coupling parameter ($\alpha$) from angle-dependent magneto-transport measurements on these samples can reveal further information about the z-confinement effect. We have estimated the phase coherence time $\tau_\phi$ as a function of the orientation of the magnetic field, which is shown in Figure 4(c). For the 5 uc sample, there is a noticeable variation in $\tau_\phi$ with angle due to the orbital contribution to the magnetoconductance, whereas the change is very small for 15 uc sample, confirming the stronger z-confinement for the 15 uc case. Figure 4(d) shows the variation of spin-orbit coupling strength, $\alpha$ with $\phi$. In the case of 5 uc, $\alpha$ decreases monotonically as $\phi$ is varied from 0 to 90°. This can be explained by an additional out-of-plane Rashba SOC term (arising from broken symmetry within the plane of the interface) in addition to an in-plane Rashba SOC contribution (arising from broken symmetry perpendicular to the interface). It is to be noted that the out-of-plane Rashba SOC is also gate voltage tunable. It is possible that the presence of atomic planes with imperfect step edges may break in-plane crystal symmetry which can create potential gradients $(\partial V / \partial x)$ and $(\partial V / \partial y)$. This combined with atomic spin-orbit coupling could create an out-of-plane Rashba SOC (where the in-plane Rashba SOC is related to $(\partial V / \partial z)$). In-plane anisotropic electrical properties predominantly due to atomic step edges have been reported in LAO/STO (001) system[40], however concrete evidence was lacking. Therefore, our study establish this experimental result. A theoretical study of the anisotropy of the two dimensional electron gas predicts a similar behavior due to broken in-plane symmetry.[41] An increase in $\alpha$ was observed experimentally at W(110) and Mo (110) upon Li-coverage, even though the out-of-plane SOC was not considered in the paper.[42] On the other hand, in the case of the 15 uc sample, the variation of $\alpha$ with $\phi$ is very small. This is because of the stronger confinement effect of the 2D electrons, which leads to a larger $(\partial V / \partial z)$ compared to $(\partial V / \partial x)$ and $(\partial V / \partial y)$ resulting in a stronger in-plane SOC and the resultant angular insensitivity of $\alpha$. A theoretical study[41] predicts larger $\alpha$ in the presence of an out-of-plane SOC and that is why 5 uc sample has larger $\alpha$ than 15 uc sample where the out-of-plane SOC is absent.

To further investigate the presence of out-of-plane SOC, in-plane magnetoconductance measurements with the applied magnetic field in the plane of the 2DEG, have been carried out at different gate voltages for the 5 uc and 15 uc samples which is shown in **Figure 5**(a) and 5(b). It is inferred that the magnetoconductance is gate tunable in the 5 uc case just like the out-of-plane magnetoconductance whereas the gate tunability of magnetoconductance is weak in the 15 uc case. This strongly suggests the non-negligible contribution of in-plane potential gradients in the case of 5 uc LAO which leads to an out-of-plane Rashba SOC which is also gate tunable. The very small gate tunability of 15 uc suggests the dominance of in-plane Rashba SOC in comparison to out-of-plane Rashba SOC.

It is clear from the above discussion that the SOC is clearly affected by the occupation of orbitals which in this case is $d_{xz}/d_{yz}$.[35] The carrier density in 5 uc sample is lower than 15 uc sample, which implies a different SOC strength between these two samples due to a difference in occupancy as well as the quantum confinement. Moreover the anisotropic band structure makes the SOC tunable via crystallography as well as by $V_G$. Theoretical calculations including strain, quantum confinement and orbital occupancy in the basic band structure is needed to fully understand the origin of SOC in this system.

## 3. Conclusions

A gate tunable in- and out-of-plane Rashba SOC has been observed in the case of 5 uc of $LaAlO_3/SrTiO_3$ (110) suggesting the interface has only a quasi-two dimensional character, whereas the sample with 15 uc of LAO shows only in-plane Rashba SOC suggesting increased 2D confinement. Thus the out-of-plane SOC can be controlled by varying LAO thickness. A spin splitting energy of 25 meV at the $LaAlO_3/SrTiO_3$ (110) interface is estimated for in-plane spins at a current density of $1.4 \times 10^4$ A/cm$^2$, which is large compared to that found in semiconductor heterostructures or the $LaAlO_3/SrTiO_3$ (100) interface, and in addition it is anisotropic. The anisotropy is understood to arise from the difference in the effective mass associated with Ti $t_{2g}$ orbitals along [001] and [1-10]. We also show with two different LAO thicknesses that strain can affect the orbital occupancy and the strength of SOC and mobility which can be modified by an applied field. Hence a detailed theoretical study on the effect of strain, z-confinement, band filling and crystal orientation on SOC may enable us to engineer the interface for spin-orbit torque device applications.

## 4. Experimental Section

Samples of LAO with 5 and 15 unit cells (uc) were grown layer by layer on Ti-terminated STO (110) substrates at an oxygen partial pressure of 1 mTorr using pulsed laser deposition. Prior to the deposition of epitaxial LAO, amorphous AlN of 50 nm thick has been deposited on the lithographically patterned sample. The interface between STO and amorphous AlN do not exhibit any conductivity and only the LAO/STO interface is conducting. Samples with several other thickness of LAO have also been prepared and characterized, but we have chosen two representative samples for the discussion here. Further details on the sample preparation and characterization are given in Ref. [20] and supplementary materials.

## Supporting Information

Supporting Information is available from the Wiley Online Library or from the author.

## Acknowledgements

We thank B. Gu, C. Nayak, S. Maekawa, Q. Zhang, and S. Yunoki for the helpful discussions. This work was supported by National Research Foundation (NRF) Singapore under the competitive research program (CRP) "Control of exotic quantum phenomena at strategic interfaces and surfaces for novel functionality by in situ synchrotron radiation" (CRP Award No. NRF-CRP8-2011-06) and "New approach to low power information storage: electric-field controlled magnetic memories" (CRP Award No. NRF-CRP10-2012-02).

[1]    R. S. Keizer, S. T. B. Goennenwein, T. M. Klapwijk, G. Miao, G. Xiao, and A. Gupta, *Nature* **2006**, 439, 825.

[2]    H. Zhang, C.-X. Liu, X.-L. Qi, X. Dai, Z. Fang, and S.-C. Zhang, *Nat. Phys.* **2009**, 5, 438.

[3]    F. Wilczek, *Nat. Phys.* **2009**, 5, 614.

[4]    S. Datta and B. Das, *Appl. Phys. Lett.* **1990**, 56, 665.

[5]    A. R. Mellnik, J. S. Lee, A. Richardella, J. L. Grab, P. J. Mintun, M. H. Fischer, A. Vaezi, A. Manchon, E. A. Kim, N. Samarth, and D. C. Ralph, *Nature* **2014**, 511, 449.

[6]    I. Mihai Miron, G. Gaudin, S. Auffret, B. Rodmacq, A. Schuhl, S. Pizzini, J. Vogel, and P. Gambardella, *Nat. Mater.* **2010**, 9, 230.

[7]    J. P. Eisenstein, H. L. Störmer, V. Narayanamurti, A. C. Gossard, and W. Wiegmann, *Phys. Rev. Lett.* **1984**, 53, 2579.

[8]    J. Nitta, T. Akazaki, H. Takayanagi, and T. Enoki, *Phys. Rev. Lett.* **1997**, 78, 1335.

[9]     K. Narayanapillai, K. Gopinadhan, X. Qiu, A. Annadi, Ariando, T. Venkatesan, and H. Yang, *Appl. Phys. Lett.* **2014**, 105, 162405.

[10]    A. D. Caviglia, M. Gabay, S. Gariglio, N. Reyren, C. Cancellieri, and J. M. Triscone, *Phys. Rev. Lett.* **2010**, 104, 126803.

[11]    M. Ben Shalom, M. Sachs, D. Rakhmilevitch, A. Palevski, and Y. Dagan, *Phys. Rev. Lett.* **2010**, 104, 126802.

[12]    A. F. Santander-Syro, F. Fortuna, C. Bareille, T. C. Rödel, G. Landolt, N. C. Plumb, J. H. Dil, and M. Radović, *Nat. Mater.* **2014**, 13, 1085.

[13]    D. A. Dikin, M. Mehta, C. W. Bark, C. M. Folkman, C. B. Eom, and V. Chandrasekhar, *Phys. Rev. Lett.* **2011**, 107, 056802.

[14]    Ariando, X. Wang, G. Baskaran, Z. Q. Liu, J. Huijben, J. B. Yi, A. Annadi, A. R. Barman, A. Rusydi, S. Dhar, Y. P. Feng, J. Ding, H. Hilgenkamp, and T. Venkatesan, *Nat. Commun.* **2011**, 2, 188.

[15]    N. Reyren, S. Thiel, A. D. Caviglia, L. F. Kourkoutis, G. Hammerl, C. Richter, C. W. Schneider, T. Kopp, A. S. Ruetschi, D. Jaccard, M. Gabay, D. A. Muller, J. M. Triscone, and J. Mannhart, *Science* **2007**, 317, 1196.

[16]    L. Li, C. Richter, J. Mannhart, and R. C. Ashoori, *Nat. Phys.* **2011**, 7, 762.

[17]    K. Michaeli, A. C. Potter, and P. A. Lee, *Phys. Rev. Lett.* **2012**, 108, 117003.

[18]    A. Brinkman, M. Huijben, M. van Zalk, J. Huijben, U. Zeitler, J. C. Maan, W. G. van der Wiel, G. Rijnders, D. H. A. Blank, and H. Hilgenkamp, *Nat. Mater.* **2007**, 6, 493.

[19]    G. Herranz, F. Sánchez, N. Dix, M. Scigaj, and J. Fontcuberta, *Sci. Rep.* **2012**, 2, 758.

[20]    A. Annadi, Q. Zhang, X. Renshaw Wang, N. Tuzla, K. Gopinadhan, W. M. Lü, A. Roy Barman, Z. Q. Liu, A. Srivastava, S. Saha, Y. L. Zhao, S. W. Zeng, S. Dhar, E. Olsson, B. Gu, S. Yunoki, S. Maekawa, H. Hilgenkamp, T. Venkatesan, and Ariando, *Nat. Commun.* **2013**, 4, 1838.

[21]    A. Joshua, S. Pecker, J. Ruhman, E. Altman, and S. Ilani, *Nat. Commun.* **2012**, 3, 1129.

[22]    C. Cancellieri, D. Fontaine, S. Gariglio, N. Reyren, A. D. Caviglia, A. Fête, S. J. Leake, S. A. Pauli, P. R. Willmott, M. Stengel, P. Ghosez, and J. M. Triscone, *Phys. Rev. Lett.* **2011**, 107, 056102.

[23]    M. Basletic, J. L. Maurice, C. Carretero, G. Herranz, O. Copie, M. Bibes, E. Jacquet, K. Bouzehouane, S. Fusil, and A. Barthelemy, *Nat. Mater.* **2008**, 7, 621.

[24]    C. W. Bark, P. Sharma, Y. Wang, S. H. Baek, S. Lee, S. Ryu, C. M. Folkman, T. R. Paudel, A. Kumar, S. V. Kalinin, A. Sokolov, E. Y. Tsymbal, M. S. Rzchowski, A. Gruverman, and C. B. Eom, *Nano Lett.* **2012**, 12, 1765.

[25]    C. Bell, S. Harashima, Y. Hikita, and H. Y. Hwang, *Appl. Phys. Lett.* **2009**, 94, 222111.

[26]    R. Pentcheva and W. E. Pickett, *Phys. Rev. Lett.* **2009**, 102, 107602.

[27]    P. D. Dresselhaus, C. M. A. Papavassiliou, R. G. Wheeler, and R. N. Sacks, *Phys. Rev. Lett.* **1992**, 68, 106.

[28]    J. Nitta, T. Akazaki, H. Takayanagi, and T. Enoki, *Phys. Rev. Lett.* **1997**, 78, 1335.

[29]    Y. Kim, R. M. Lutchyn, and C. Nayak, *Phys. Rev. B* **2013**, 87, 245121.

[30]    S. Maekawa and H. Fukuyama, *J. Phys. Soc. Jpn* **1981**, 50, 2516.

[31]    L. F. Mattheiss, *Phys. Rev. B* **1972**, 6, 4740.

[32]    R. J. Elliott, *Phys. Rev.* **1954**, 96, 266.

[33]    M. I. D'yakonov and V. I. Perel', *Sov. Phys. Solid State* **1972**, 13, 3023.

[34]    Z. Zhong, A. Tóth, and K. Held, *Phys. Rev. B* **2013**, 87, 161102.

[35]    Z. Wang, Z. Zhong, X. Hao, S. Gerhold, B. Stöger, M. Schmid, J. Sánchez-Barriga, A. Varykhalov, C. Franchini, K. Held, and U. Diebold, *PNAS* **2014**, 111, 3933.

[36]    J. Nitta, T. Akazaki, H. Takayanagi, and T. Enoki, *Phys. Rev. Lett.* **1997**, 78, 1335.

[37]    T. Koga, J. Nitta, T. Akazaki, and H. Takayanagi, *Phys. Rev. Lett.* **2002**, 89, 046801.

[38]    D. Grundler, *Phys. Rev. Lett.* **2000**, 84, 6074.

[39]    M. Salluzzo, J. C. Cezar, N. B. Brookes, V. Bisogni, G. M. De Luca, C. Richter, S. Thiel, J. Mannhart, M. Huijben, A. Brinkman, G. Rijnders, and G. Ghiringhelli, *Phys. Rev. Lett.* **2009**, 102, 166804.

[40]    P. Brinks, W. Siemons, J. E. Kleibeuker, G. Koster, G. Rijnders, and M. Huijben, *Appl. Phy. Lett.* **2011**, 98, 242904.

[41]    J. Premper, M. Trautmann, J. Henk, and P. Bruno, *Phys. Rev. B* **2007**, 76, 073310.

[42]    E. Rotenberg, J. W. Chung, and S. D. Kevan, *Phys. Rev. Lett.* **1999**, 82, 4066.

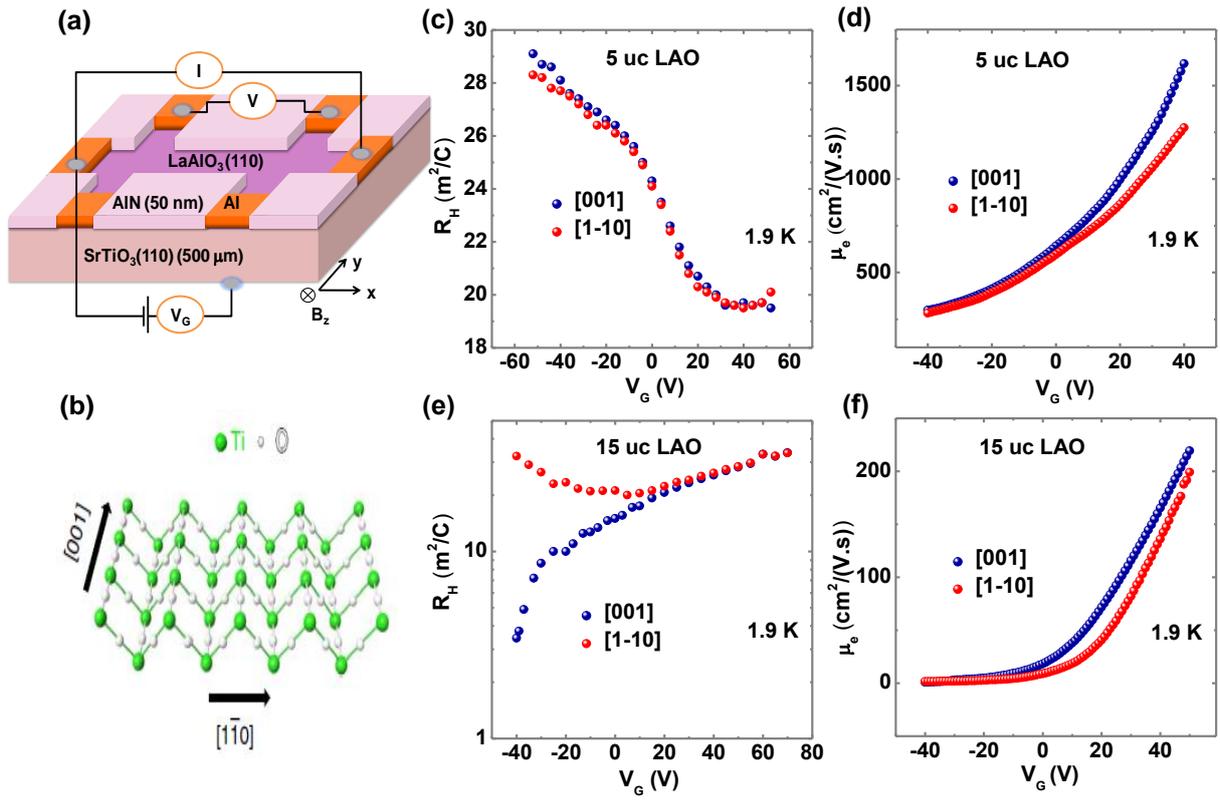

**Figure 1**. (a) Schematic diagram of the measurement geometry. (b) Schematic representation of the Ti and O arrangement along [1-10] and [001] of the Ti-terminated SrTiO$_3$ (110) surface. (c) Hall coefficient (R$_H$) and (d) mobility ($\mu_e$) as a function of back gate voltage (V$_G$) along [001] and [1-10] for 5 uc of LAO at a temperature of 1.9 K. (e) Hall coefficient (R$_H$) and (f) mobility ($\mu_e$) as a function of back gate voltage (V$_G$) along [001] and [1-10] for 15 uc of LAO.

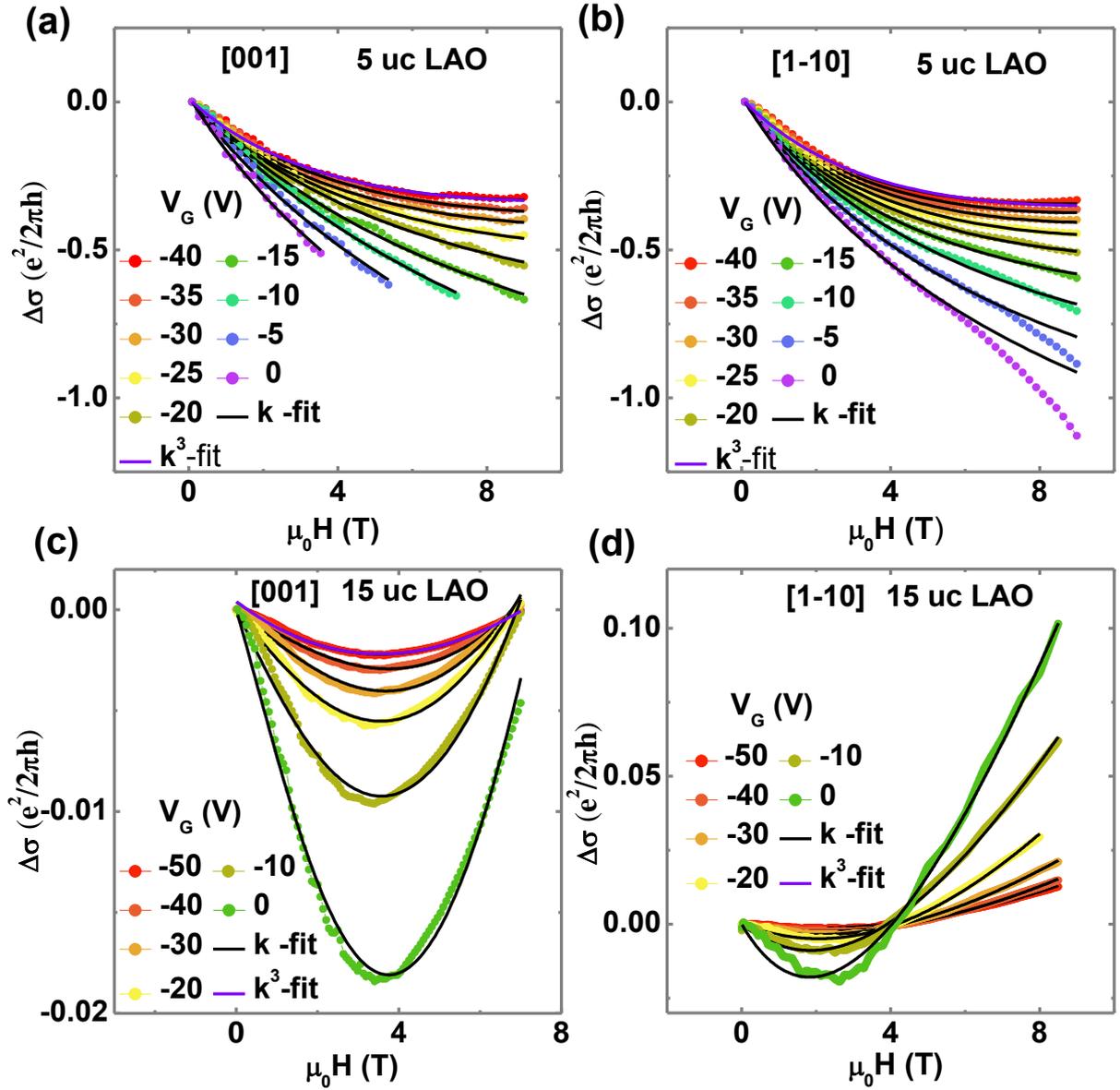

**Figure 2.** Magnetoconductance ($\Delta\sigma$) vs. applied field H as a function of back gate voltage ($V_G$) along (a) [001] and (b) [1-10] for 5 uc of LAO and along (c) [001] and (d) [1-10] for 15 uc of LAO measured at 1.9 K. Fits to the Kim-Lutchyn-Nayak theory are shown by the solid lines assuming either a linear ($k$) or a cubic ($k^3$) Rashba effect at the interface.

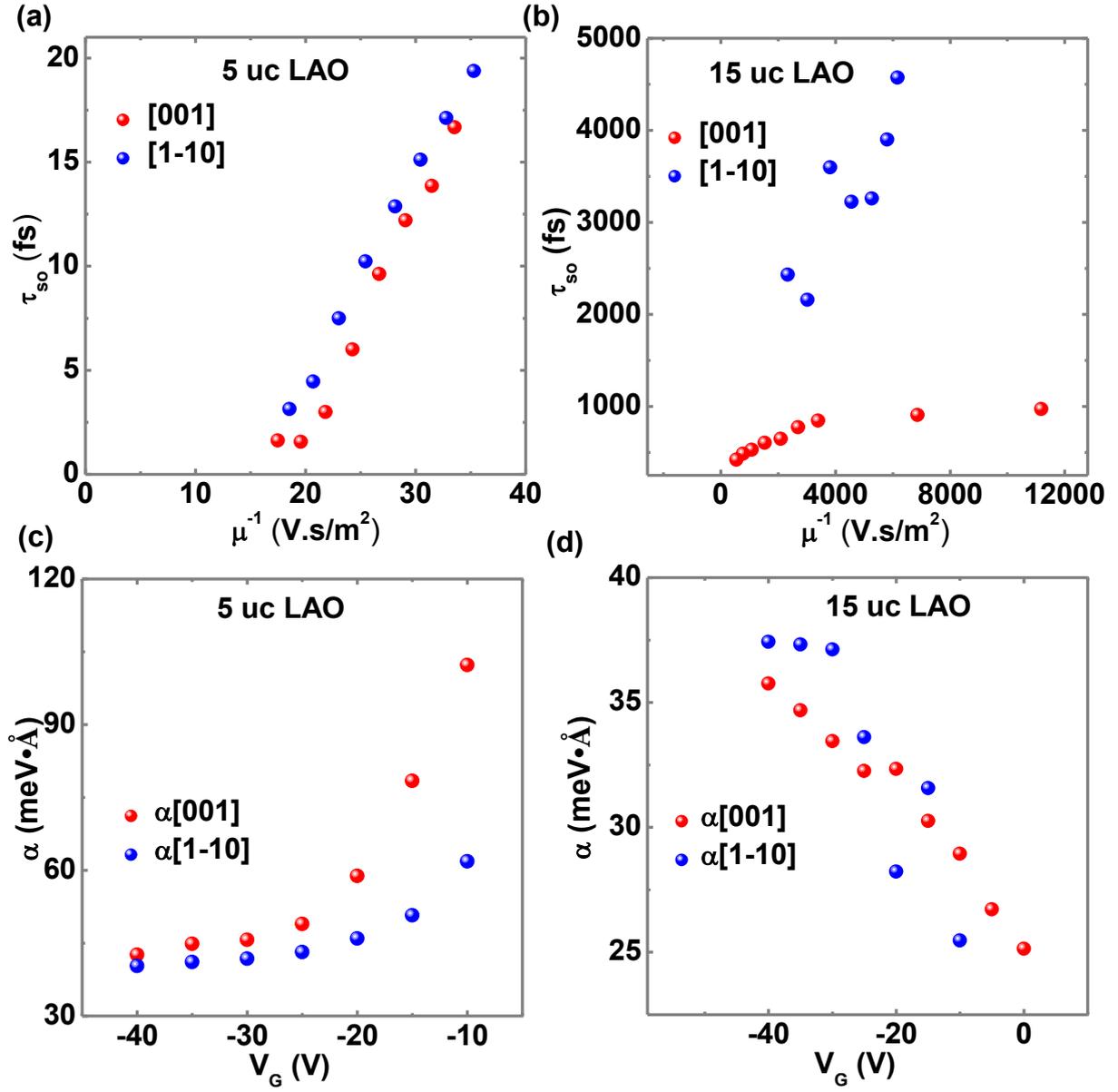

**Figure 3**. Estimated spin-orbit relaxation time ($\tau_{so}$) as a function of inverse mobility ($\mu$) along [001] and [1-10] for (a) 5 uc of LAO and for (b) 15 uc of LAO sample. Estimated Rashba coupling constant ($\alpha$) as a function of gate voltage ($V_G$) along [001] and [1-10] for (c) 5 uc LAO and for (d) 15 uc LAO sample.

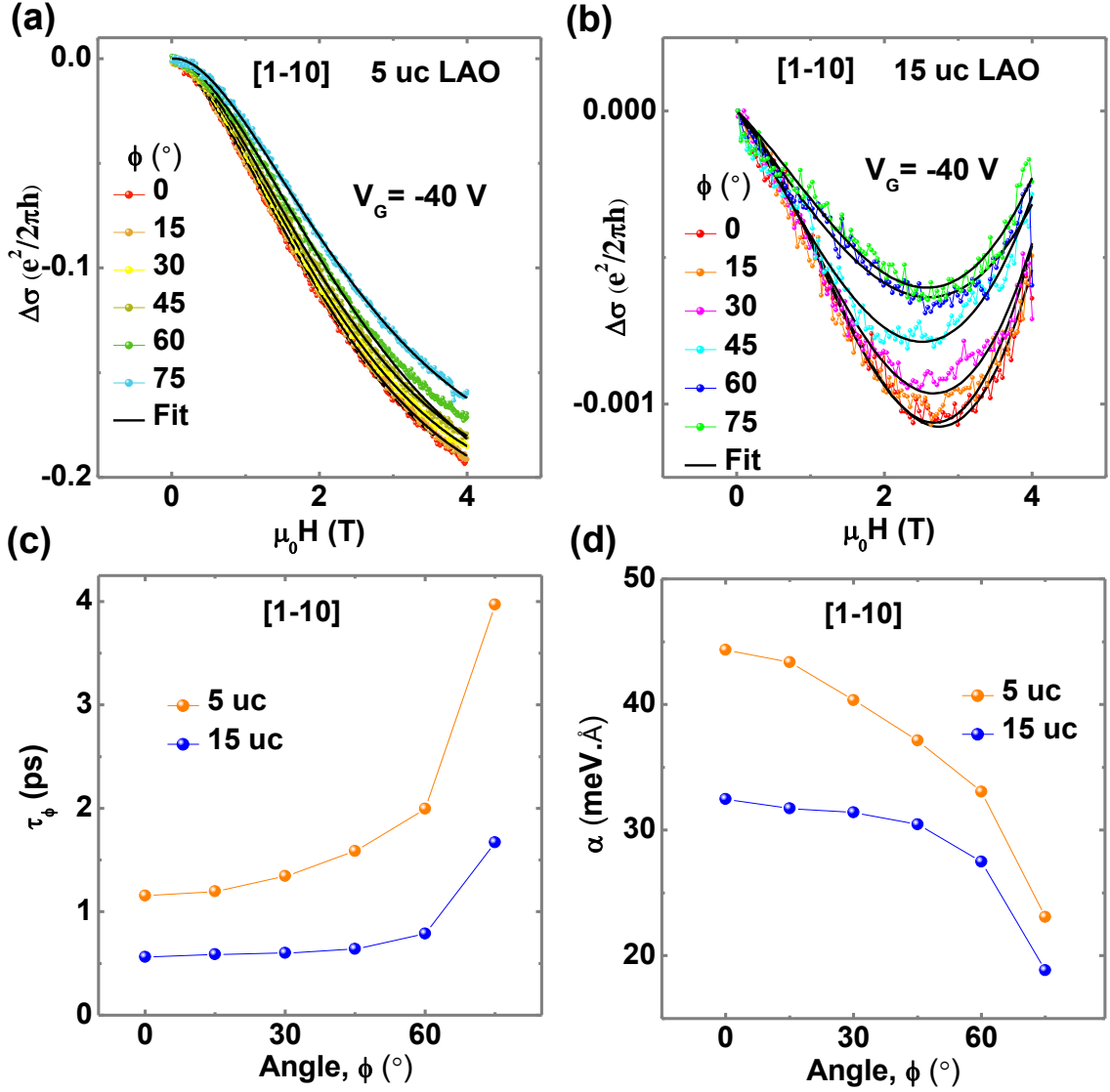

**Figure 4**. Magnetoconductance ($\Delta\sigma$) at different angles of the magnetic field H ranging from out-of-plane ($\phi = 0°$) to in-plane ($\phi = 90°$) for (a) 5 uc of LAO and (b) 15 uc of LAO sample. The current and magnetic field are along [1-10] where the magnetic field is in the plane ($\phi = 90°$). (c) Estimated Rashba coupling constant ($\alpha$) as a function of the angle of the magnetic field for 5 and 15 uc of LAO sample. (d) Inelastic phase coherence time ($\tau_\phi$) as a function of the angle of the magnetic field for 5 and 15 uc of LAO sample.

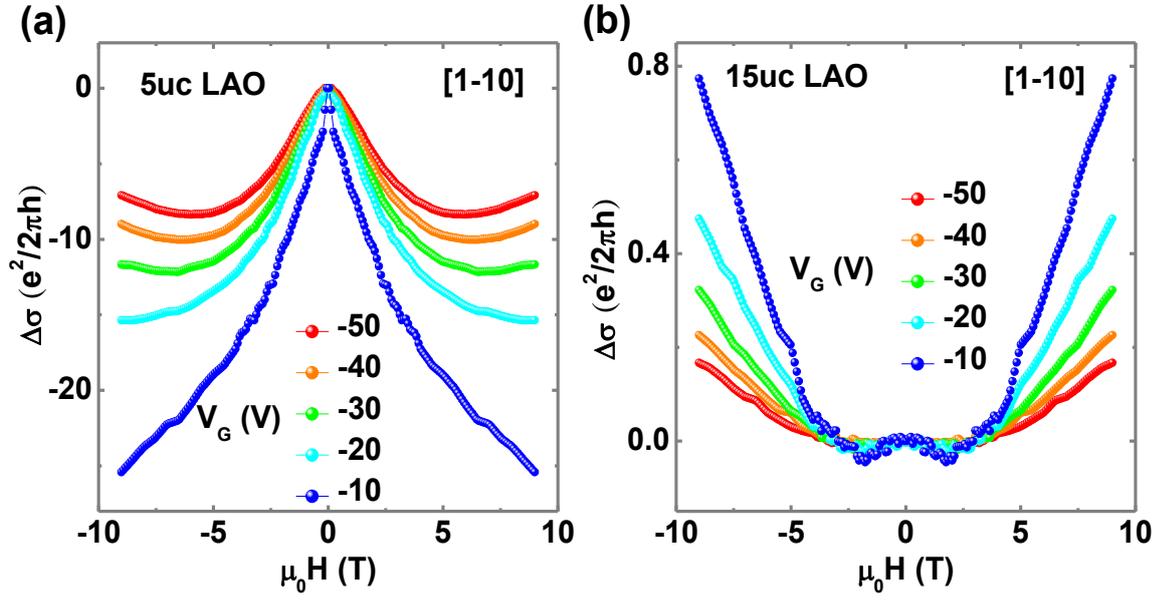

**Figure 5**. In-plane magnetoconductance (Δσ) as a function of gate voltage (V$_G$) for (a) 5 uc and (b) 15 uc LAO. It is clear that the magnetoconductance is gate tunable in the case of 5 uc sample just like the out-of-plane magnetoconductance whereas the gate tunability of magnetoconductance is weak in the case of 15 uc sample.